\documentstyle[11pt]{article}
\begin{document}
\begin{center}
{\bf NEW PUBLICATION\\
\vspace{0.5cm}

ASTROPHYSICAL PLASMAS AND FLUIDS\\
\vspace{0.5cm}

By\\
\vspace{0.5cm}

Vinod Krishan\\
Indian Institute of Astrophysics, Bangalore, India\\
\vspace{0.5cm}

Astrophysics and space science Library 235}\\
\end{center}
\vspace{0.5cm}

\noindent
{\bf Abstract}:
This book is a valuable introduction to astrophyscial plasmas and fluids
for graduate students of astronomy preparing either for a research career
in the field or just aspiring to achieve a decent degree of familiarity
with 99\% of the cosmos.
\vspace{0.2cm}

\noindent
The contents provide a true representation of the phenomenal diversity of
dominant roles that plasmas and fluids play in the near and far reaches of
the universe.  The breadth of coverage of basic physical processes is a
particularly attractive feature of this text book.  By first using the
Liouville equation to derive the kinetic, the two-fluid and single-fluid,
descriptions of a plasma and a fluid, and then demonstrating the use of
these descriptions for specific situations in the rest of the book, the
author has probably chosen the most efficient way of handling this large
technical subject.  The two major astrophysical issues, fluid or plasma
configurations and their radiative signatures, figure prominently
througout the book.  The problems are designed to give the reader a feel
for the quantititative properties of celestial objects.
\vspace{0.3cm}

\noindent
{\bf Contents}:
\vspace{0.3cm}

\noindent
1. Plasma - The Universal State of Matter, 2. Statistical Description of a
Many-Body System.  3. Particle and Fluid Motions in Gravitational and
Electromagnetic Fields. 4. Magnetohydrodynamics of Conducting Fluids. 5.
Two-Fluid Descriptions of plasma.  6. Kinetic Description of Plasmas. 7.
Nonconducting Astrophysical Fluids. 8. Physical Constants. 9.
Astrophyscial Quantities. 10. Differential Operators. 11. Characteristic
Numbers for Fluids. 12. Acknowledgement for Figures. Index.
\vspace{0.3cm}

\noindent
`This book is an excellent introduction and fills an important niche. 
This book has an excellent chance to attract a large audience.  An added
bonus is the wonderful final hapter on neutral fluid dynamics.  The
emphasis on turbulence is an excellent corrective to the usual laminar
approach to fluid dynamics.  The tone of this text is unique.  The
organization is excellent and the coverage very appropriate for an
introductory text'.
\vspace{0.3cm}

\noindent
Paul J.Wiita, Georgia State University, USA.
\vspace{0.3cm}

\noindent
November 1998, 372 pp.\\
Hardbound, ISBN 0-7923-5312-9\\
NLG 295, 00/USD 159.00/GBP 100.00\\
\vspace{0.3cm}

\noindent
November 1998, 372 pp.\\
Paperback, ISBN 0-7923-5490-7\\
NLG 130.00/USD 70.00/GBP 45.00\\
\vspace{0.3cm}

\noindent
ORDER FORM:\\
\vspace{0.3cm}

\noindent
Please send me Astrophysical Plasmas and Fluids by Vinod Krishan:
\vspace{0.3cm}

\noindent
-Copy(ies) of Hardbound, ISBN 0-7923-5312-9\\
NLG 295.00/USD 159.00/GBP  100.00\\
\vspace{0.3cm}

\noindent
-Copy(ies) of Paperback, ISBN 0-7923-5490-7\\
NLG 130.00/USD 70.00/GBP 45.00\\
\vspace{0.3cm}

\noindent
Payment enclosed to the amount of...
\vspace{0.3cm}

\noindent
Please invoice me \hspace{2cm}    Please charge my credit card\\
\vspace{0.3cm}

\noindent
Am.Ex \hspace{1cm} Visa \hspace{1cm} Diners club \hspace{1cm} Mastercard  \hspace{1cm}
Eurocard
\vspace{0.3cm}

\noindent
Access
\vspace{0.3cm}

\noindent
Name of Card Holder:
\vspace{0.3cm}

\noindent
Card No.:            \hspace{5cm}      Expiry Date:
\vspace{0.3cm}

\noindent
Delivery address:
\vspace{0.3cm}

\noindent
Name:
\vspace{0.3cm}

\noindent
Address:
\vspace{1cm}

\noindent
Date:    \hspace{5cm}    Signature:\\
\vspace{0.3cm}

\noindent
European VAT Registration Number:
\vspace{0.3cm}

\noindent
To be sent to your supplier of:
\vspace{0.3cm}

\noindent
KLUWER Academic Publishers
Order Department, P.O.Box 322
3300 AH Dordrecht, The Netherlands 
Fax: +31-78-6546474
Tel: +31-78-6392392
Internet E-mail: orderdept@skap.nl
\vspace{0.3cm}

\noindent
Orders from individuals accompanied by payment or authorization to charge
a creditcard account will ensure prompt delivery.  Postage and handling on
all such orders, delivered by surface mail, will be absorbed by the
Publisher.  Orders from outside Europe will be sent by airmail, for which
the customer will be charged extra.  Payment will be accepted in any
convertible currency.  Please check the rate of exchange at your bank.  US
Dollar prices apply to deliveries in USA, Canada and Mexico only. Prices
are subject to change without notice.  All prices are exclusive of value
added tax (VAT).  Customers in the Netherlands please add 6% VAT.
Customers from other countires in the European Community please fill in
the VAT number of your institute/company in the appropriate space on the
order form; or add 6% VAT to the total amount (customers from the UK are
not charged VAT).
\vspace{0.3cm}

\begin{center}
{\bf KLUWER ACADEMIC PUBLISHERS}
\end{center}
\end{document}